\documentclass[twocolumn,prx,aps,showpacs,amsmath,amssymb]{revtex4-1}

\usepackage{graphicx}
\usepackage{dcolumn}
\usepackage{bm}

\begin{document}

\title{Majorana tunneling entropy}

\author{Sergey Smirnov}
\affiliation{Institut f\"ur Theoretische Physik, Universit\"at Regensburg,
  D-93040 Regensburg, Germany}

\date{\today}

\begin{abstract}
In thermodynamics a macroscopic state of a system results from a number of its
microscopic states. This number is given by the exponent of the system's
entropy $\exp(S)$. In non-interacting systems with discrete energy spectra,
such as large scale quantum dots, $S$ as a function of the temperature has
usually a plateau shape with integer values of $\exp(S)$ on these
plateaus. Plateaus with non-integer values of $\exp(S)$ are fundamentally
forbidden and would be thermodynamically infeasible. Here we investigate the
entropy of a non-interacting quantum dot coupled via tunneling to normal
metals with continuum spectra as well as to topological superconductors. We
show that the entropy may have non-integer plateaus if the topological
superconductors support weakly overlapping Majorana bound states. This brings
a fundamental change in the thermodynamics of the quantum dot whose specific
heat $c_V$ acquires low temperature Majorana peaks which should be absent
according to the conventional thermodynamics. We also provide a fundamental
thermodynamic understanding of the transport properties, such as the linear
conductance. In general our results show that the thermodynamics of systems
coupled to Majorana modes represents a fundamental physical interest with
diverse applications depending on versatility of possible coupling
mechanisms.
\end{abstract}

\pacs{74.45.+c, 74.25.Bt, 65.40.gd, 74.78.Na, 74.55.+v}

\maketitle

\section{Introduction}\label{Intro}
Majorana fermions \cite{Majorana_1937}, particles identified with their own
antiparticles, have recently received a considerable attention in condensed
matter physics after the seminal Kitaev's proposal \cite{Kitaev_2001} of a
spinless chain model which could be realized using a topological
superconducting state especially in mesoscopic setups
\cite{Alicea_2012,Flensberg_2012}.

In particular, topological insulators \cite{Hasan_2010,Qi_2011} in combination
with $s$-wave superconductors are natural candidates for a practical
implementation \cite{Fu_2008,Fu_2009} of the Kitaev's model because the edge
states in topological insulators provide a single pair of the Fermi points.

Another practical implementation \cite {Lutchyn_2010,Oreg_2010} of the
Kitaev's model is based on spin-orbit coupled one-dimensional quantum wires
placed in an external magnetic field freezing out one spin component. When
this effectively "spinless" system is proximity coupled to an $s$-wave
superconductor, the one-dimensional wire becomes a topological superconductor
implementing the Kitaev's model.

As soon as the the Majorana bound states are created there appears the
question how to experimentally detect them. Since these are zero energy modes
of a mesoscopic system, it is natural to try to detect them by applying a bias
voltage to this mesoscopic system and measuring the current flowing through
it. The differential conductance then should have a maximum at zero bias. This
has been done in Ref. \cite{Mourik_2012} and signatures of the presence of the
Majorana fermions in the system have been reported.
\begin{figure}
\includegraphics[width=8.0 cm]{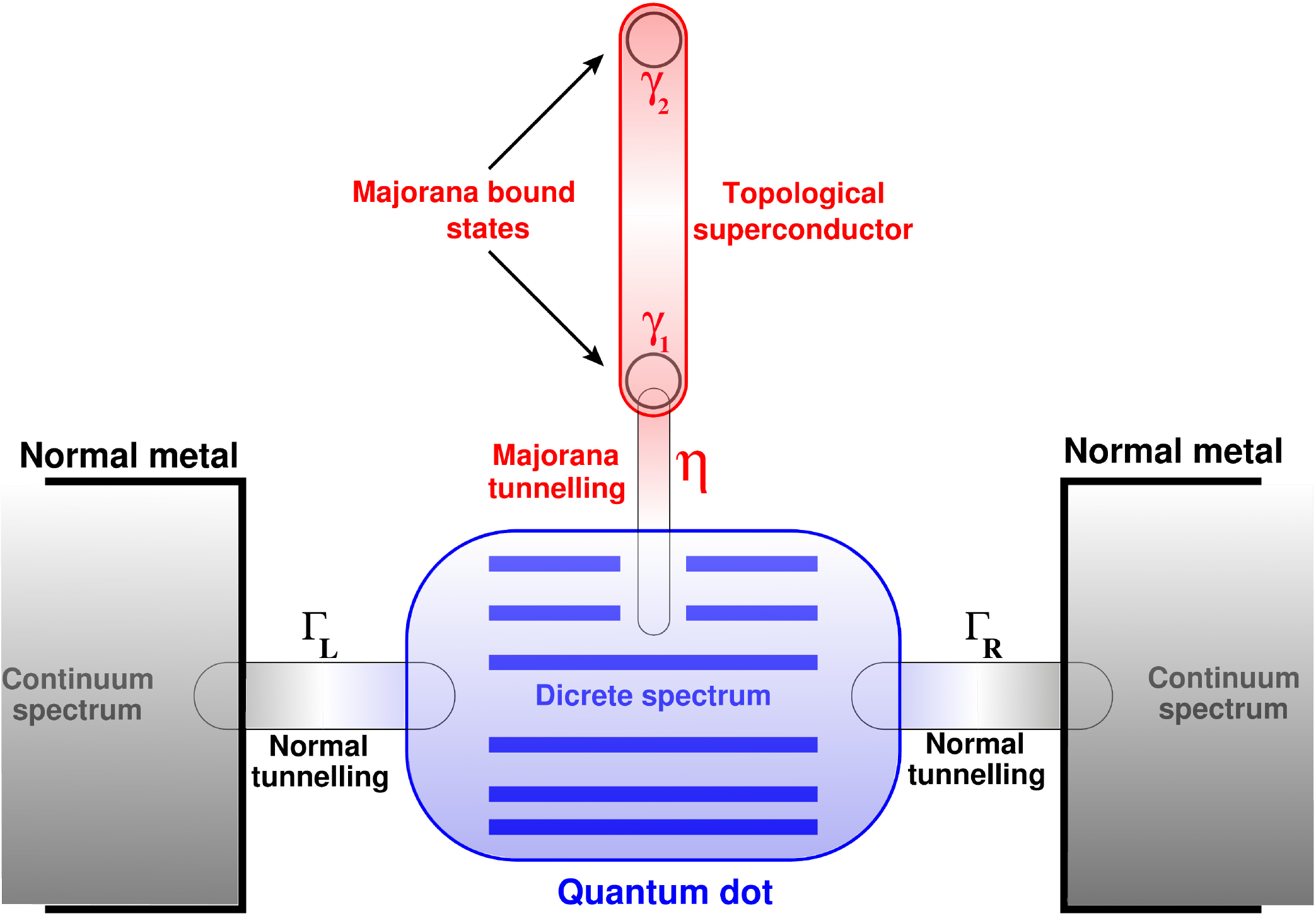}
\caption{\label{figure_1} Schematic picture of the setup. The quantum dot here
  represents a non-interacting system with a discrete energy spectrum. It is
  coupled via tunneling junctions to left and right normal metallic contacts
  (in general the number of normal contacts may be arbitrary) whose energy
  spectra are continuum. The strength of this tunneling coupling is
  characterized by the energies $\Gamma_\text{L,R}$ (see the text for the
  details). Besides, the quantum dot is coupled to a topological
  superconductor whose ends support two Majorana bound states described by the
  second quantized operators $\gamma_{1,2}$. This tunneling coupling is
  realized via only one Majorana mode $\gamma_1$ and is characterized by the
  energy $\eta$ (see the text for the details).}
\end{figure}

An alternative possibility to detect Majorana fermions is to couple the system
supporting Majorana bound states to a normal system via tunneling. The
Majorana fermions are then entangled with normal fermions changing drastically
the behavior of the initially normal system. This approach was used in
Ref. \cite{Vernek_2014} where the normal system was a non-interacting quantum
dot as well as in Refs. \cite{Cheng_2014,Liu_2015,Tijerina_2015} where the
normal system was a quantum dot with interactions. In both interacting and
non-interacting cases the emphasis was made on the transport properties of the
quantum dots coupled to topological superconductors. The differential
conductance was calculated and the zero bias Majorana peak was found. In the
interacting case an interplay between the Majorana and Kondo physics was
investigated calculating in addition to the current also the shot noise which
is the current-current correlation.

Below we adhere to that alternative strategy but instead of the transport
properties of a system coupled to a topological superconductor (or to several
topological superconductors) we focus on the thermodynamics of this system and
calculate its entropy $S$ and specific heat $c_V$. We demonstrate that 1) $S$
as a function of the temperature has a plateau shape with integer values of
$\exp(S)$ on these plateaus in the absence of the topological superconductor
or when the two Majorana modes within each topological superconductor strongly
overlap but 2) acquires additional plateaus with non-integer values of
$\exp(S)$ when only one Majorana mode from each topological superconductor is
entangled via tunneling mechanisms with the normal fermions in the quantum
dot. These additional plateaus excite in the temperature dependence of $c_V$
3) additional Majorana peaks being its first low temperature peaks.

Crucially, as it will become obvious from the text below, this alternative
approach is very attractive from the point of view of experimental
thermodynamic signatures of Majorana fermions because the quantum dot can be
viewed just as a model of a low energy spectrum of an arbitrary finite
(discrete energy spectrum) macroscopic system coupled via tunneling mechanisms
to topological superconductors. Therefore, our results are quite general and
simple to detect in modern experiments. This is in contrast to setups
\cite{Tsvelick_1984,Affleck_1991,Emery_1992} where the Majorana thermodynamics
requires specific Kondo spin-flip correlations with the Hamiltonian parameters
fine tuned which is difficult to implement in real experiments as well as in
contrast to setups with more ingredients such as the Josephson vortex dynamics
\cite{Hou_2012} which complicates the experimental observation of the Majorana
physics and in contrast to setups with only continuum spectra
\cite{Affleck_2013} where the entropy as a function of the temperature does
not have a plateau shape and, therefore, it is impossible in a realistic
experiment (always performed at finite temperatures) to disentangle the
Majorana and normal fermions contributions to the entropy.

The paper is organized as follows. In Section \ref{MT} we discuss a setup
with Majorana tunneling processes involved and calculate its
entropy. The thermodynamic results are presented and analyzed in Section
\ref{RD}. With Section \ref{Concl} we conclude the paper.
\section{Majorana thermodynamics}\label{MT}
The system under investigation is shown in Fig. \ref{figure_1}. The quantum
dot represents a system with a discrete energy spectrum. In the present study
we consider a non-interacting quantum dot and, therefore, its Hamiltonian has
the form $\hat{H}_\text{QD}=\sum_{\alpha,\alpha'}h_{\alpha\alpha'} d_\alpha^\dagger
d_{\alpha'}$, where $\alpha$ is a single-particle index. The quantum dot is
coupled to a number (two in the figure but can be arbitrary) of normal
metallic contacts with continuum spectra via normal tunneling
$\hat{H}_\text{CD}=\sum_{l=\text{L,R}}\sum_{k_l,\alpha}c_{lk_l}^\dagger
T_{lk_l,\alpha}d_\alpha+\text{H.c.}$, where $k_l$ is the
set of quantum numbers characterizing the contact with the number $l$. The
contacts are assumed to be non-interacting,
$\hat{H}_\text{C}=\sum_{l=\text{L,R}}\sum_{k_l}\epsilon_{lk_l}c_{lk_l}^\dagger
c_{lk_l}$. In addition to the normal tunneling the quantum dot is coupled via
another tunneling mechanism to a topological superconductor characterized by
the two Majorana modes $\gamma_j$, $\gamma_j^\dagger =\gamma_j$, $\gamma_j^2
=1$, $j=1,2$. This Majorana tunneling \cite{Flensberg_2010} involves only one
Majorana mode $\gamma_1$, and has the form
$\hat{H}_\text{DM}=\sqrt{2}\sum_\alpha \eta^*_\alpha
d_\alpha^\dagger\gamma_1+\text{H.c.}$. The effective low energy Hamiltonian of
the topological superconductor is given in terms of the two Majorana modes,
$\hat{H}_\text{TS}=i\xi\gamma_2\gamma_1$, where $\xi$ is the energy
characterizing the overlap strength \cite{Alicea_2012} of the two Majorana
bound states. This implies, similar to many other works on Majorana physics
(see, {\it e.g.}, Refs. \cite{Cheng_2014} and \cite{Flensberg_2010}), that we
consider energy scales (temperature, tunneling rates, etc.) much smaller than
the superconducting gap $\Delta$.

To find the entropy of the quantum dot we construct the field integral in
imaginary time \cite{Altland_2010} for the total partition function
$Z_\text{tot}$, transform it into a skew-symmetric form, integrate out all the
fermionic degrees of freedom, compute the Pfaffian of the matrix of the
imaginary time action and obtain the quantum dot partition function as
$Z=Z_\text{tot}/Z_C$, where $Z_C$ is the partition function of the normal
metallic contacts.

For the thermodynamic potential $\Omega=-\ln Z/\beta$ we then obtain
\begin{equation}
\Omega=\Omega_\text{TS}+\frac{1}{\beta}\sum_{n\ge 0}\ln\text{det}[\mathcal{G}_{i\alpha,i'\alpha'}(\omega_n)],
\label{TP_non_diag}
\end{equation}
where $\Omega_\text{TS}$ is the thermodynamic potential of the isolated
topological superconductor, $\beta$ is the inverse temperature, $\beta\equiv
1/T$ (we use the energy units for the temperature $T$, {\it i.e.},
$k_\text{B}=1$) and $\mathcal{G}_{i\alpha,i'\alpha'}(\omega_n)$ are the
Fourier transforms of the quantum dot imaginary time Green's functions,
$\mathcal{G}(i\alpha\tau|i'\alpha'\tau')\equiv\langle\text{T}\,d_{i\alpha}(\tau)d_{i'\alpha'}(\tau')\rangle$
(here $i=p,h$ and $d_{p\alpha}(\tau)\equiv d_{\alpha}^\dagger(\tau)$,
$d_{h\alpha}(\tau)\equiv d_{\alpha}(\tau)$ and the angular brackets stand for
the thermal average), taken at the discrete Matsubara fermionic frequencies
$\omega_n\equiv\pi(2n+1)/\hbar\beta$.

Choosing the the single-particle states $|\alpha\rangle$ such that
$h_{\alpha\alpha'}=\delta_{\alpha\alpha'}\epsilon_\alpha$ the imaginary time
Green's functions become diagonal and Eq. (\ref{TP_non_diag}) reduces to
\begin{equation}
\begin{split}
&\Omega=\Omega_\text{TS}+\frac{1}{2\beta}\sum_{\alpha,n}\ln[\mathcal{G}_{\alpha\,hp}^*(\omega_n)\mathcal{G}_{\alpha\,hp}(\omega_n)-\\
&-\mathcal{G}_{\alpha\,pp}^*(\omega_n)\mathcal{G}_{\alpha\,pp}(\omega_n)].
\end{split}
\label{TP_diag}
\end{equation}
From Eq. (\ref{TP_diag}) we find the entropy $S=-\partial\Omega/\partial T$,
\begin{equation}
\begin{split}
&S=\ln\biggl[\cosh\biggl(\frac{\xi}{2T}\biggl)\biggl]-\frac{\xi}{2T}\tanh\biggl(\frac{\xi}{2T}\biggl)+\ln(2)+\\
&+\frac{1}{16\pi iT^2}\sum_\alpha
\int_{-\infty}^\infty
d\epsilon\frac{\epsilon}{\cosh^2\bigl(\frac{\epsilon}{2T}\bigl)}\ln[G_\alpha(\epsilon)],\\
\end{split}
\label{En}
\end{equation}
where
\begin{equation}
G_\alpha(\epsilon)\equiv\frac{G_{\alpha\,hp}^A(-\epsilon)G_{\alpha\,hp}^R(\epsilon)-
G_{\alpha\,hh}^A(-\epsilon)G_{\alpha\,pp}^R(\epsilon)}
{G_{\alpha\,hp}^R(-\epsilon)G_{\alpha\,hp}^A(\epsilon)-
G_{\alpha\,hh}^R(-\epsilon)G_{\alpha\,pp}^A(\epsilon)}.
\label{Galpha}
\end{equation}
In Eq. (\ref{Galpha}) $G^{R,A}_{\alpha\,ii'}(\epsilon)$ are the Fourier
transforms of the diagonal elements of the quantum dot real time retarded and
advanced Green's functions, $iG^{R,A}(i\alpha t|i'\alpha'
t')\equiv\pm\Theta(\pm t\mp
t')\langle\{d_{i\alpha}(t),d_{i'\alpha'}(t')\}\rangle$.

We now apply the general Eqs. (\ref{En}) and (\ref{Galpha}) to the case when
the quantum dot has a single energy level $\epsilon_d$. The normal fermions in
the quantum dot are characterized by the spin index,
$|\alpha\rangle=|\sigma\rangle$, and, as in Ref. \cite{Liu_2015}, we assume
that only one spin component couples to the normal metallic contacts and
topological superconductor ($\eta_\sigma$ is zero for one spin component and
is equal to $\eta$ for the opposite one). The single-particle index and
summation over it are then irrelevant in Eq. (\ref{En}).

To find the retarded and advanced Green's functions we construct the Keldysh
field integral \cite{Altland_2010}, perform the Keldysh rotation to get the
the retarded-advanced structure of the Keldysh action, integrate out the
fermionic degrees of freedom of the normal metallic contacts and topological
superconductor and take the retarded and advanced elements in the inverse
matrix of the Keldysh effective action.
\begin{figure}
\includegraphics[width=8.0 cm]{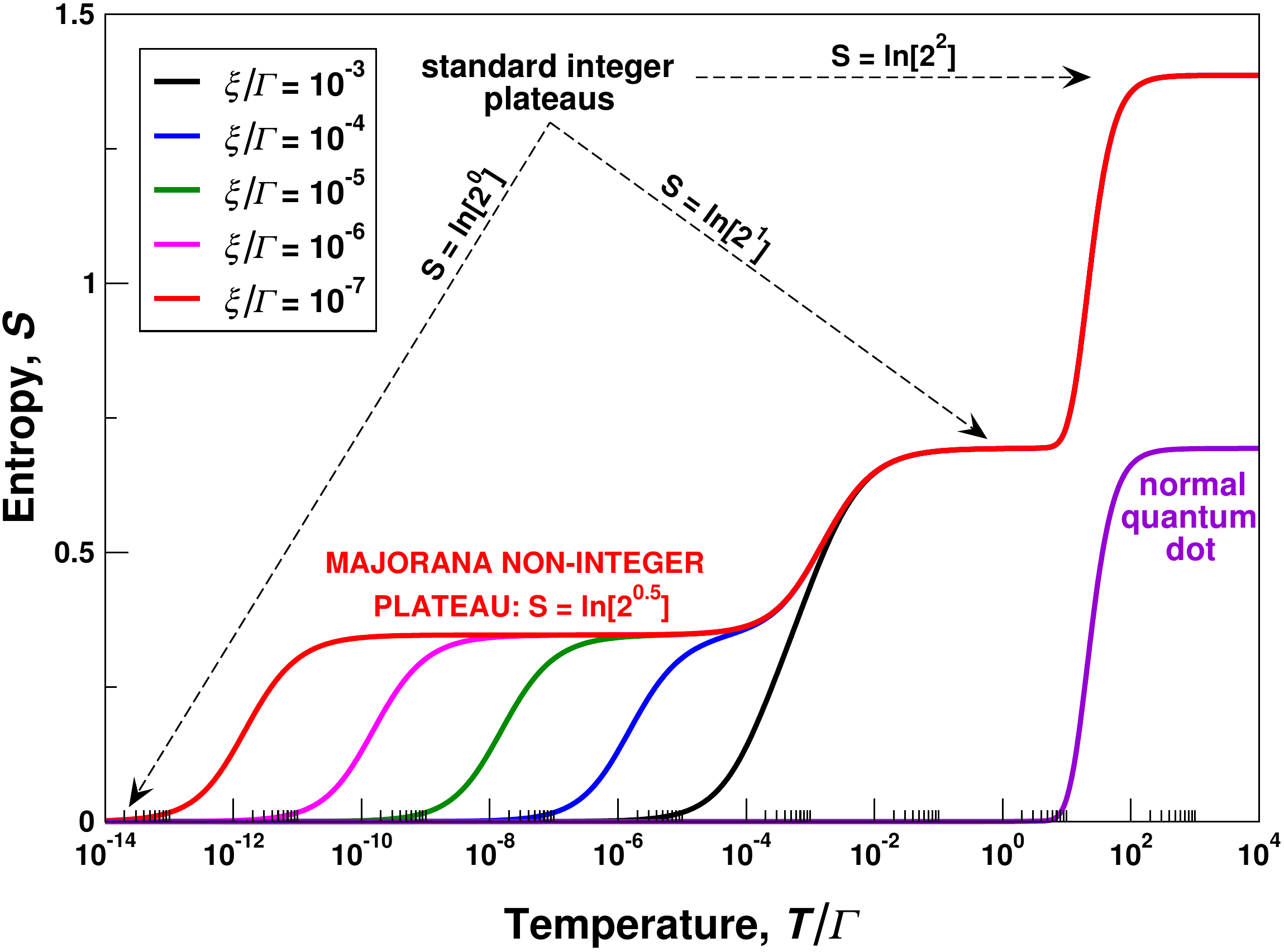}
\caption{\label{figure_2} Tunneling entropy $S$ as a function of the
  temperature $T/\Gamma$ for different values of the overlap energy
  $\xi/\Gamma$ of the Majorana bound states. The other parameters are
  $\epsilon_d/\Gamma=-50.0$, $\eta/\Gamma=2.0$. The purple curve shows the
  tunneling entropy for a normal quantum dot that is the one which is not
  coupled to any topological superconductor. As expected for a system with a
  discrete spectrum, $S$ has plateaus. When the overlap of the two Majorana
  modes is strong, there are only standard plateaus with integer values of
  $\exp(S)$ similar to normal quantum dots (purple curve). However, when the
  overlap of the Majorana bound states is weak, there appears additional
  plateau with a non-integer value of $\exp(S)$ equal to $\sqrt{2}$.}
\end{figure}
\begin{figure}
\includegraphics[width=8.0 cm]{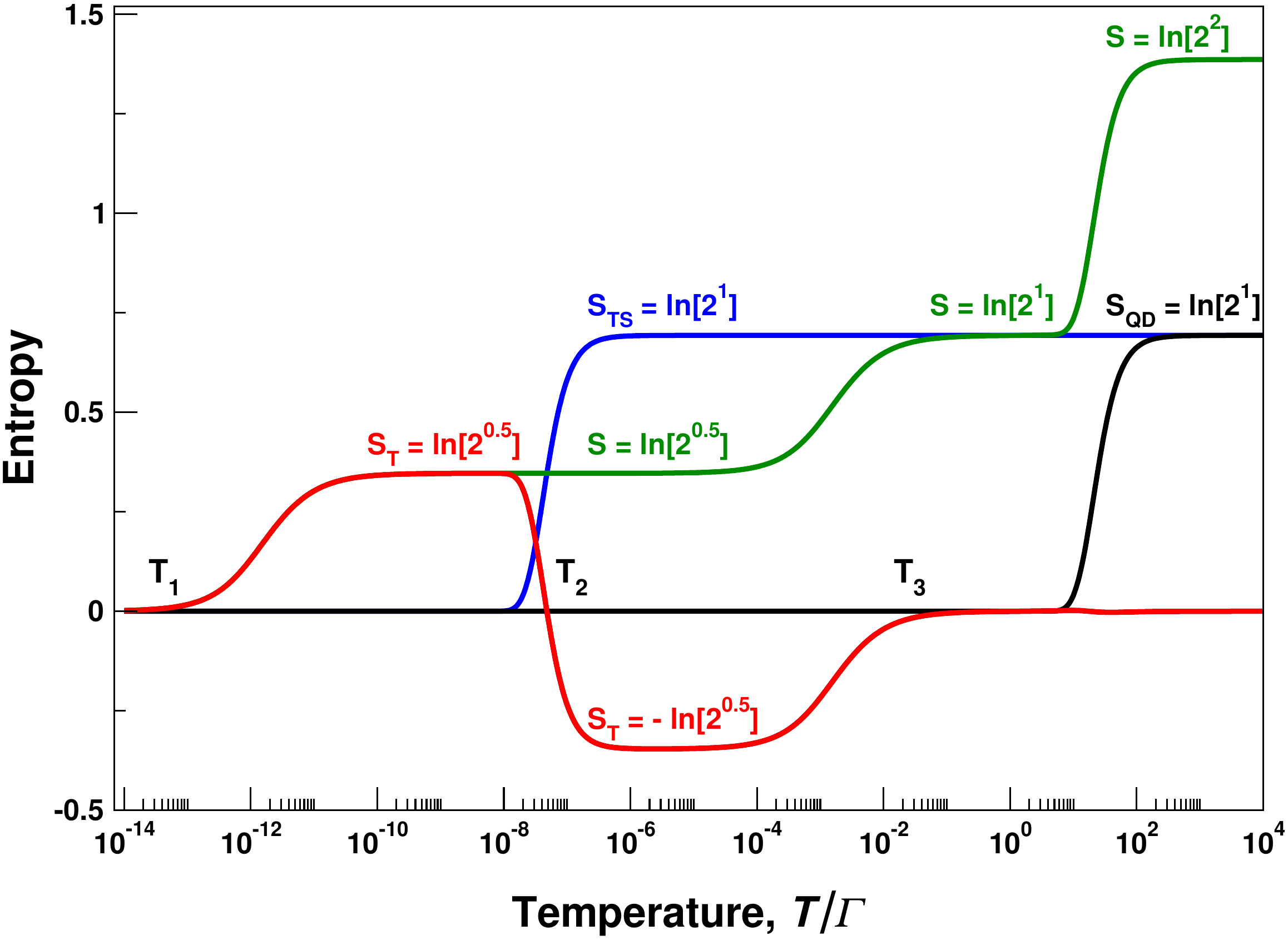}
\caption{\label{figure_3} Tunneling entropy $S$ (green curve) is the sum of
  three contributions: the entropy of the quantum dot $S_\text{QD}$ (black
  curve), the entropy of the topological superconductor $S_\text{TS}$ (blue
  curve) and the entropy of the tunneling interaction between the quantum dot
  and topological superconductor $S_\text{T}$ (red curve). The characteristic
  energies $T_1$, $T_2$ and $T_3$ (see the main text) indicate the temperature
  ranges where $S_\text{T}$ has the positive and negative plateaus with the
  absolute value $\ln[2^{0.5}]$. Here $\xi/\Gamma=10^{-7}$. The other
  parameters have the same values as in Fig. \ref{figure_2}.}
\end{figure}

For simplicity we assume that all the normal contacts are described by the
same quantum numbers $k=\{r,\sigma\}$ ($r$ is the orbital part and $\sigma$ is
the spin part) and are characterized by a constant density of states
$\nu_\text{C}$. Additionally, to simplify the calculations we assume that
$T_{lr\sigma,\sigma'}=\delta_{\sigma\sigma'}T_{l\sigma}$ ($T_{l\sigma}$ is
zero for the same spin component for which $\eta_\sigma$ is zero as mentioned
above). We then obtain:
\begin{equation}
\begin{split}
&G_{hp}^R(\epsilon)=\frac{2\hbar\{-4|\eta|^2\epsilon-(\xi^2-\epsilon^2)[i\Gamma+2(\epsilon_d+\epsilon)]\}}{f(\epsilon)},\\
&G_{pp}^R(\epsilon)=\frac{-8\hbar\eta^2\epsilon}{f(\epsilon)},\quad
G_{hh}^R(\epsilon)=\frac{-8\hbar(\eta^*)^2\epsilon}{f(\epsilon)},\\
&f(\epsilon)=(\Gamma^2+4\epsilon_d^2)\xi^2-4i\Gamma\epsilon(2|\eta|^2+\xi^2)-\\
&-\epsilon^2[16|\eta|^2+\Gamma^2+4(\epsilon_d^2+\xi^2)]+4i\Gamma\epsilon^3+4\epsilon^4
\end{split}
\label{GRA}
\end{equation}
and $G_{hp}^A(\epsilon)=[G_{hp}^R(\epsilon)]^*$,
$G_{pp}^A(\epsilon)=[G_{hh}^R(\epsilon)]^*$,
$G_{hh}^A(\epsilon)=[G_{pp}^R(\epsilon)]^*$. In Eq. (\ref{GRA})
$\Gamma\equiv\Gamma_\text{L}+\Gamma_\text{R}$ and
$\Gamma_l\equiv\pi\nu_C|T_l|^2$.

Using Eqs. (\ref{En}), (\ref{Galpha}) and (\ref{GRA}) one can calculate the
entropy of the quantum dot:
\begin{equation}
\begin{split}
&S=\ln\biggl[\cosh\biggl(\frac{\xi}{2T}\biggl)\biggl]-\frac{\xi}{2T}\tanh\biggl(\frac{\xi}{2T}\biggl)+\ln(2)+\\
&+\frac{1}{8\pi T^2}\int_{-\infty}^\infty
d\epsilon\frac{\epsilon\,\phi(\epsilon)}{\cosh^2\bigl(\frac{\epsilon}{2T}\bigl)},
\end{split}
\label{En_fin}
\end{equation}
where $\phi(\epsilon)$ is the phase of the following complex function:
\begin{equation}
G_{hp}^A(-\epsilon)G_{hp}^R(\epsilon)-G_{hh}^A(-\epsilon)G_{pp}^R(\epsilon)=\rho(\epsilon)e^{i\phi(\epsilon)}.
\label{Phase}
\end{equation}
\section{Results and discussion}\label{RD}
As it is known from statistical physics
\cite{Landau_V}, a macroscopic state of a system is a collective result of a
certain number of its microscopic states. This number is provided by the
system's entropy $S$ and is equal to $\exp(S)$. When the temperature is
increased, more and more high energy states will be involved in the
macroscopic state and the entropy will grow. If the system's spectrum is
discrete and the distance between the levels is large enough, then in certain
temperature intervals the entropy will not change, {\it i.e.}, its temperature
dependence will have plateaus on which $\exp(S)$ is integer showing how many
microscopic states are involved in the temperature interval of a given
plateau and on the last, the highest, plateau the quantity $\exp(S)$ will
provide an integer number equal to the dimensionality of the Hilbert space of
the system.
\begin{figure}
\includegraphics[width=8.0 cm]{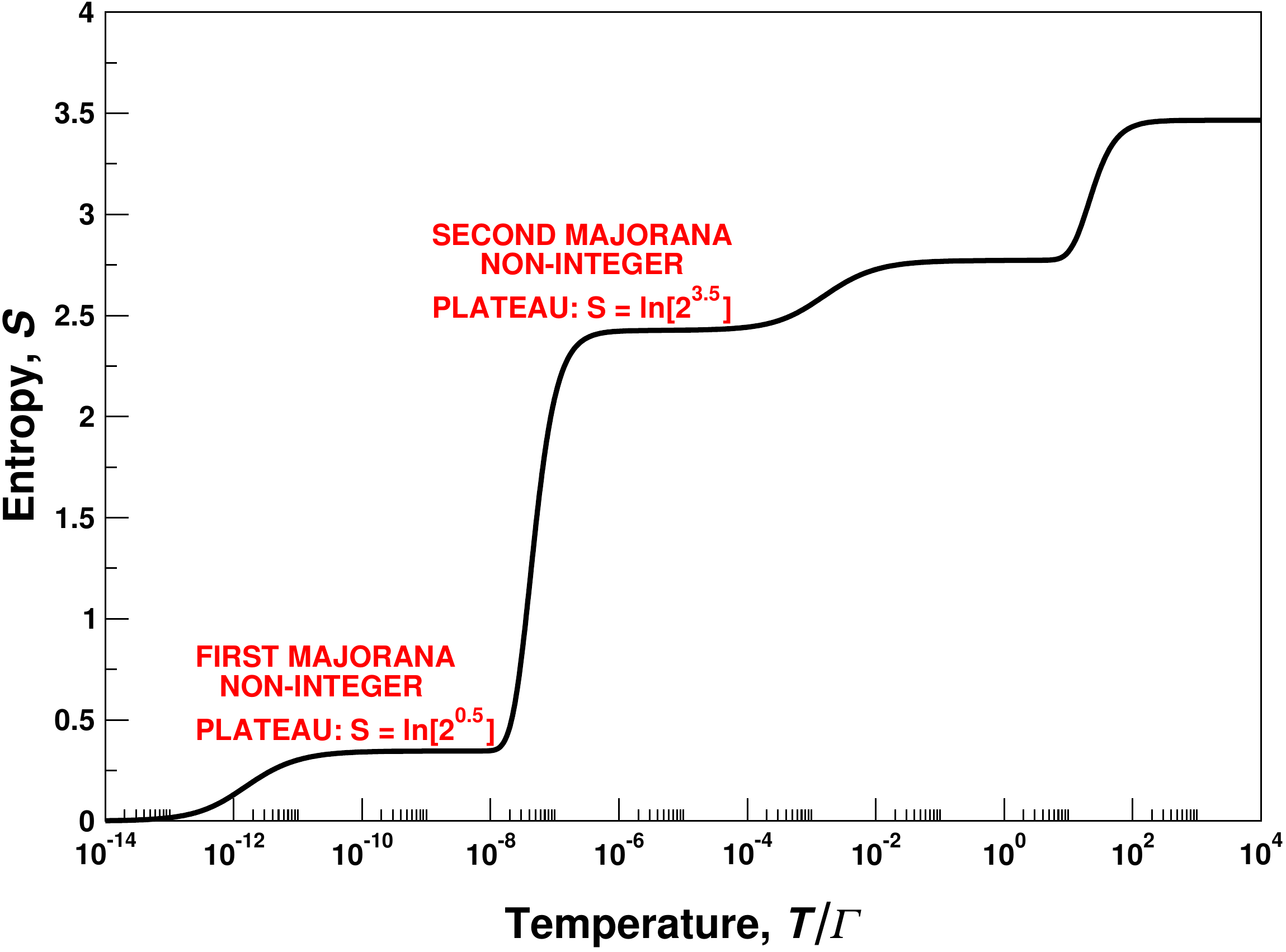}
\caption{\label{figure_4} The tunneling entropy $S$ as a function of the
  temperature $T/\Gamma$ for the case $N_\text{TS}=4$. The overlap energies of
  the topological superconductors are the same, $\xi/\Gamma=10^{-7}$. The
  other parameters have the same values as in Fig. \ref{figure_2} where now
  $\eta$ should be understood as the total tunneling strength between the
  quantum dot and topological superconductors. As expected, the tunneling
  entropy has two plateaus with non-integer values of $\exp(S)$ equal to
  $2^{0.5}$ and $2^{3.5}$.}
\end{figure}

However this standard picture described above changes drastically as soon as
the system's normal fermions are entangled with Majorana fermions via a
certain mechanism. In the present case this entanglement happens via the
tunneling of the strength $\eta$ between the quantum dot and topological
superconductor. As shown in Fig. \ref{figure_2}, in the absence of the
Majorana fermions (the lowest purple curve) the quantum dot entropy has two
standard plateaus with integer values of $\exp(S)$ equal to $1$ and $2$
corresponding to the two possible states of the quantum dot with one or zero
electrons. Similarly, as one can see in Fig. \ref{figure_2}, when the quantum
dot is coupled to a topological superconductor whose Majorana modes strongly
overlap (the black curve, $\xi/\Gamma=10^{-3}$) the entropy has plateaus only
with integer values of $\exp(S)$. The role of the Majorana fermions in this
case is just in the doubling of the dimensionality of the Hilbert space and in
the formation of the plateaus with even values of $\exp(S)$. However, when
$\xi/\Gamma$ gets smaller and the Majorana bound states overlap weakly there
appears a plateau with a non-integer value of $\exp(S)$ which is equal to
$\sqrt{2}$. This Majorana plateau represents a signature that the Majorana
modes are well separated and strongly entangled with the normal fermions in
the quantum dot whose macroscopic state at low temperatures represents now a
collective result of the Majorana microscopic states.

To better understand the relevant physics and to see its non-trivial Majorana
content the tunneling entropy $S$ might be conceived of three contributions,
$S=S_\text{QD}+S_\text{TS}+S_\text{T}$ coming from the quantum dot,
$S_\text{QD}$, from the topological superconductor, $S_\text{TS}$, and from
the tunneling interaction between the quantum dot and topological
superconductor, $S_\text{T}$. There are three characteristic energy scales,
$T_1\sim\xi^2[\xi+|\epsilon_d|\exp(-\eta/|\epsilon_d|)]/\eta^2$, $T_2\sim\xi$
and $T_3\sim\eta^2/2|\epsilon_d|$ shown in Fig. \ref{figure_3}. When $\xi$
decreases, tunneling events involve essentially one Majorana fermion. On one
side, during tunneling events the Dirac fermions of the quantum dot transform
into the Majorana fermion and vice versa. Such tunneling processes lead to an
effective  reduction of the number of the fermionic degrees of freedom. As a
result, $S_\text{T}$ is negative in the temperature range from $T_2$ to $T_3$
(red curve in Fig. \ref{figure_3}) leading to a decrease of $S$ in this
temperatures range from $S=\ln(2)$ (the value it would have had without the
tunneling) down to $S=\ln(2)/2$. On the other side, when $\xi$ is small, the
multiple tunneling processes between the
\begin{figure}
\includegraphics[width=8.0 cm]{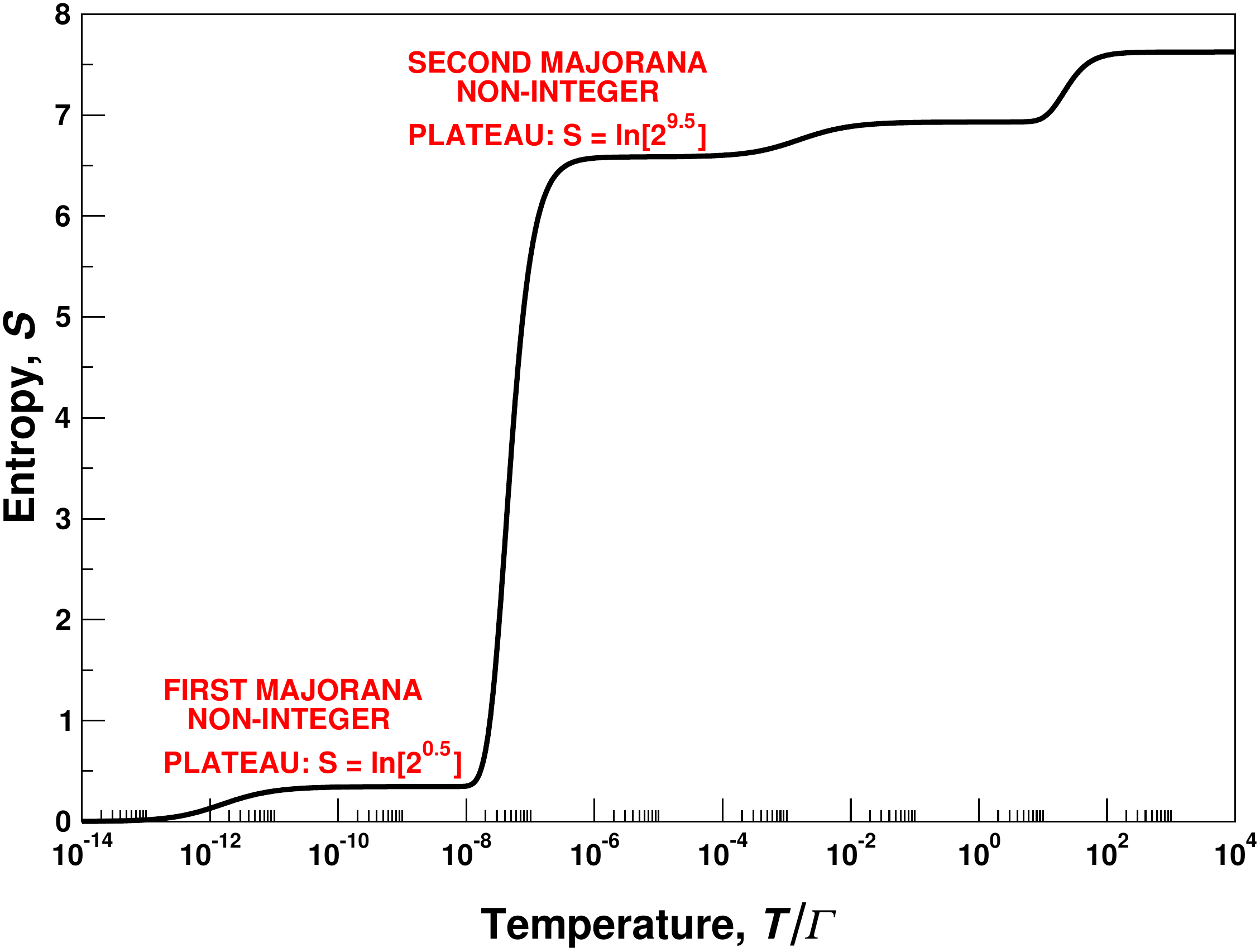}
\caption{\label{figure_5} The tunneling entropy $S$ as a function of the
  temperature $T/\Gamma$ for the case $N_\text{TS}=10$. The overlap energies
  of the topological superconductors are the same, $\xi/\Gamma=10^{-7}$. The
  other parameters have the same values as in Fig. \ref{figure_2} where now
  $\eta$ should be understood as the total tunneling strength between the
  quantum dot and topological superconductors. As expected, the tunneling
  entropy has two plateaus with non-integer values of $\exp(S)$ equal to
  $2^{0.5}$ and $2^{9.5}$.}
\end{figure}
quantum dot and topological superconductor give rise to a quantum state where
the Dirac fermions of the quantum dot are hybridized with the single Majorana
fermion of the topological superconductor in such a way that this new quantum
state has a fractional number of degrees of freedom and is responsible for the
formation of the entropy plateau $S=\ln(2)/2$ in the temperature range from
$T_1$ to $T_2$ (red curve in Fig. \ref{figure_3}). While it would be
reasonable to expect that for this hybridized state the number of degrees of
freedom should be between $2$ (Dirac fermion) and $\sqrt{2}$ (Majorana
fermion), it is surprising that in the present case it is equal to
$\sqrt{2}$. In other words, from the point of view of the number of degrees of
freedom the hybridized state could have been interpreted as a pure single
Majorana fermion. However, what is more important and highly non-trivial is
the fact that the both processes, the decrease of the entropy in the
temperature range between $T_2$ and $T_3$ and its increase in the temperature
range between $T_1$ and $T_2$, happen in such a coherent way that there
develops a single plateau of the tunneling entropy $S=\ln(2)/2$ in the
temperature range from $T_1$ to $T_3$. The necessary condition for the
validity of our analytical estimate made above for the orders of $T_1$, $T_2$
and $T_3$ as well as for the formation of a plateau with a non-integer value
of $\exp(S)$ is obviously $T_1\ll T_2\ll T_3$ which is the case for small
$\xi$ or/and large $\eta$.
\begin{figure}
\includegraphics[width=8.0 cm]{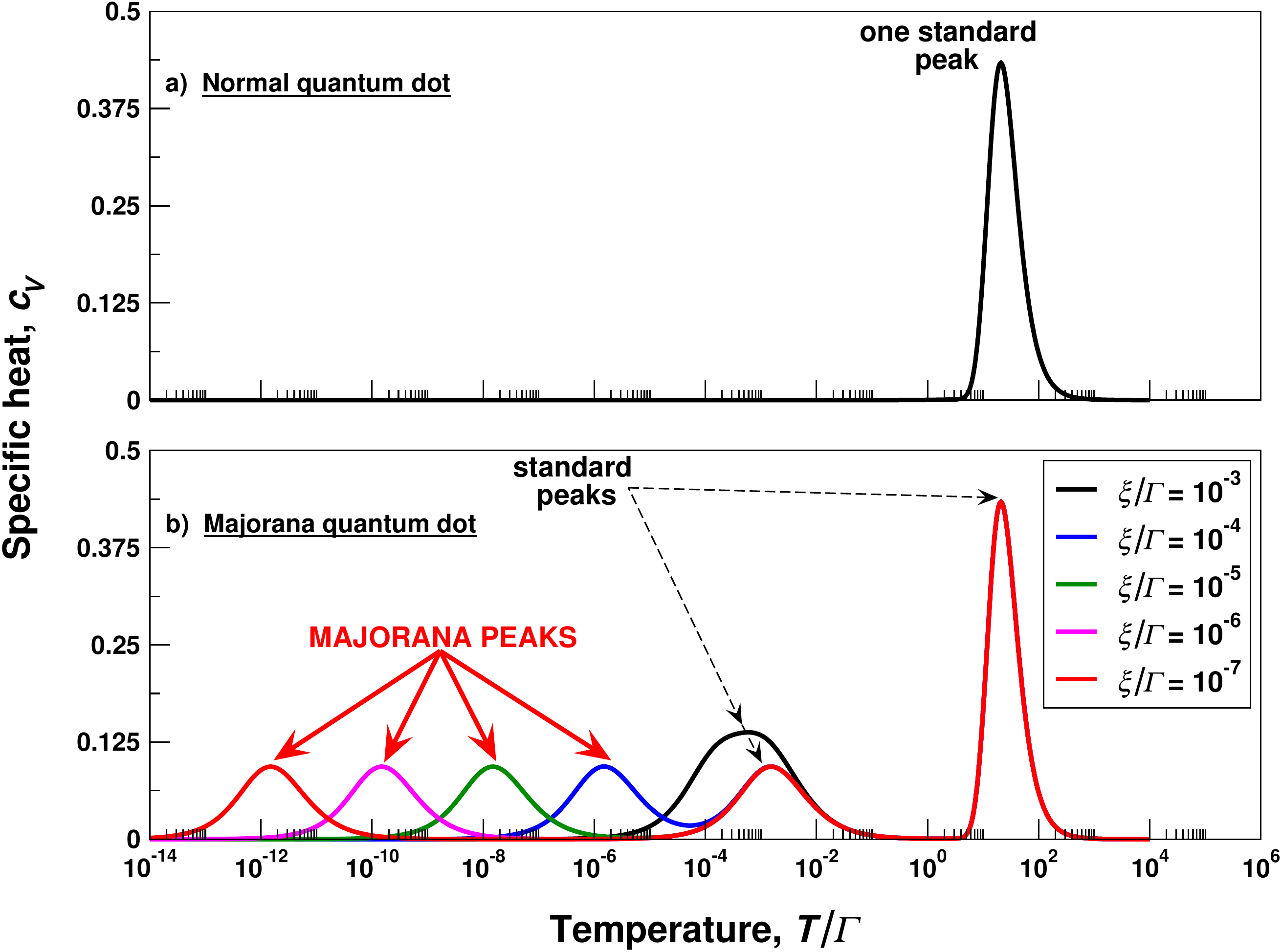}
\caption{\label{figure_6} Specific heat $c_V$ as a function of the temperature
  $T/\Gamma$. Part a) shows the specific heat for a normal quantum dot while
  Part b) shows it for a Majorana quantum dot with different values of the
  overlap energy $\xi/\Gamma$ of the Majorana bound states. The other
  parameters are as in Fig. \ref{figure_2}. The plateau shape of the tunneling
  entropy of a system with a discrete spectrum gives rise to a set of peaks in
  the specific heat. As the overlap of the Majorana bound states gets weaker,
  in addition to standard peaks there develops the Majorana leftmost peak
  which becomes the first low temperature peak of the specific heat.}
\end{figure}

From the discussion above it is obvious that the situation with a single
plateau on which $\exp(S)$ takes a non-integer value is exceptional because it
takes place only when the quantum dot is coupled to just one topological
superconductor. Indeed, one might consider a situation when the quantum dot is
coupled to $N_\text{TS}>1$ topological superconductors. This is experimentally
relevant because the quantum dot with a single level $\epsilon_d$ represents
just a model of the low energy spectrum of a large macroscopic system
whose size is much larger than the relevant size of the topological
superconductors and, therefore, this macroscopic system may easily be coupled
to many topological superconductors. Since the Majorana modes in different
topological superconductors are independent (in contrast to the Majorana modes
in the same topological superconductor), their contribution to the tunneling
entropy does not depend on the parity of $N_\text{TS}$ and, therefore,
plateaus with non-integer values of $\exp(S)$ are expected also for any
$N_\text{TS}>1$. In the case when the overlap energies of all the topological
superconductors are approximately the same it is obvious that there will be
two Majorana plateaus with $S/\ln(2)=1/2$ (between $T_1$ and $T_2$) and
$S/\ln(2)=N_\text{TS}-1/2$ (between $T_2$ and $T_3$). When $N_\text{TS}=1$,
$N_\text{TS}-1/2=1/2$ and there is only one Majorana plateau. However,
$N_\text{TS}-1/2>1/2$ for $N_\text{TS}>1$. The latter case is shown in
Figs. \ref{figure_4} and \ref{figure_5} for $N_\text{TS}=4$ and
$N_\text{TS}=10$, respectively.
\begin{figure}
\includegraphics[width=8.0 cm]{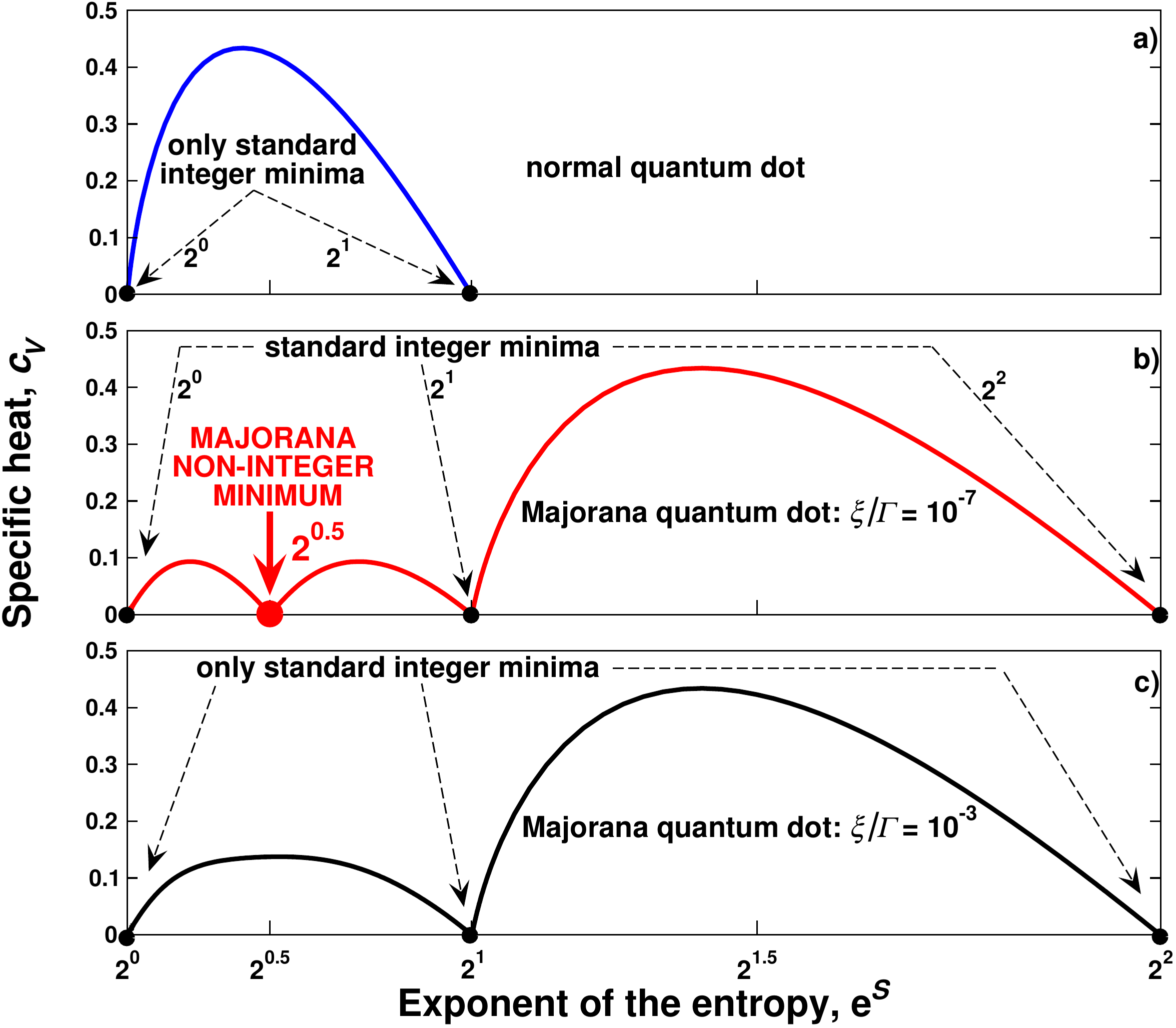}
\caption{\label{figure_7} Specific heat $c_V$ as a function of the exponent of
  the tunneling entropy, $\exp(S)$. In this representation the specific heat
  has the standard minima at integer values. Part a) shows that for a normal
  quantum dot these minima are located at $2^0$ and $2^1$. However, in a
  quantum dot coupled to a topological superconductor in addition to the
  standard minima at $2^0$, $2^1$ and $2^2$ there appears an additional
  Majorana non-integer minimum at $2^{0.5}$ as shown in Part b) for weakly
  overlapping Majorana modes, $\xi/\Gamma=10^{-7}$. When the two Majorana
  bound states strongly overlap the Majorana non-integer minimum disappears as
  shown in Part c) for $\xi/\Gamma=10^{-3}$. The other parameters are as in
  Fig. \ref{figure_2}.}
\end{figure}

The negative contribution to the entropy and the formation of the hybridized
state discussed above arise from the tunneling interaction and represent a
general mechanism which will take place in other normal systems (not
necessarily quantum dots) having tunneling contacts with Majorana systems. The
thermodynamic analysis above leads to the following general result. Since in
equilibrium the entropy $S$ is maximal, we conclude that the decrease of the
entropy $S$ induced by the tunneling interaction in the temperature range
between $T_2$ and $T_3$ has a fundamental consequence that tunneling contacts
between normal systems and Majorana systems would eventually vanish in this
temperature range and their longstanding existence would be possible only due
to an external work or a metastable state (which could have a very long
lifetime).
\begin{figure}
\includegraphics[width=8.0 cm]{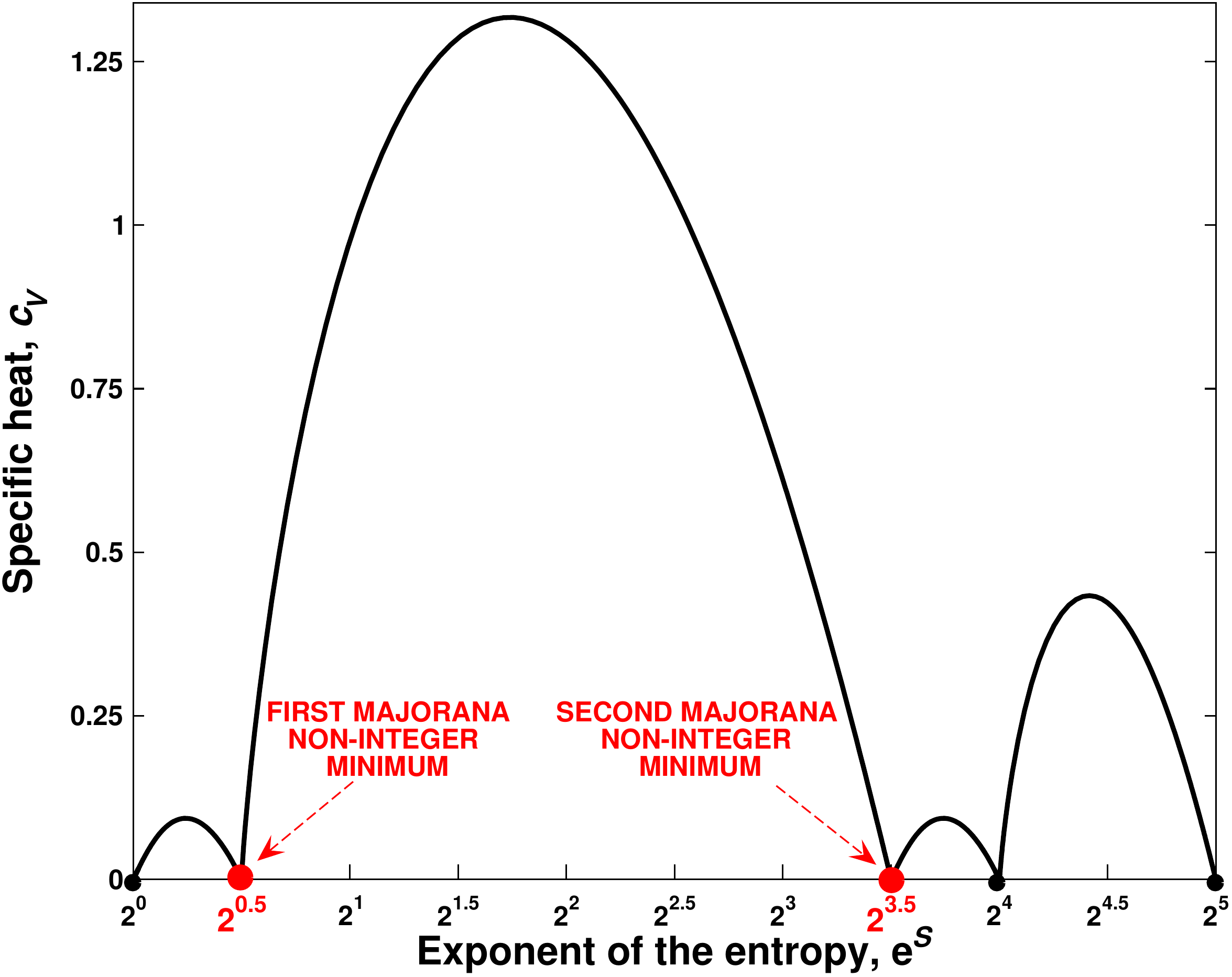}
\caption{\label{figure_8} Specific heat $c_V$ as a function of the exponent of
  the tunneling entropy, $\exp(S)$, for the case $N_\text{TS}=4$,
  $\xi/\Gamma=10^{-7}$. The other parameters have the same values as in
  Fig. \ref{figure_2} where now $\eta$ should be understood as the total
  tunneling strength between the quantum dot and topological
  superconductors. As expected, in addition to the standard minima at $2^0$,
  $2^4$ and $2^5$ there appear two additional Majorana non-integer minima at
  $2^{0.5}$ and $2^{3.5}$.}
\end{figure}
\begin{figure}
\includegraphics[width=8.0 cm]{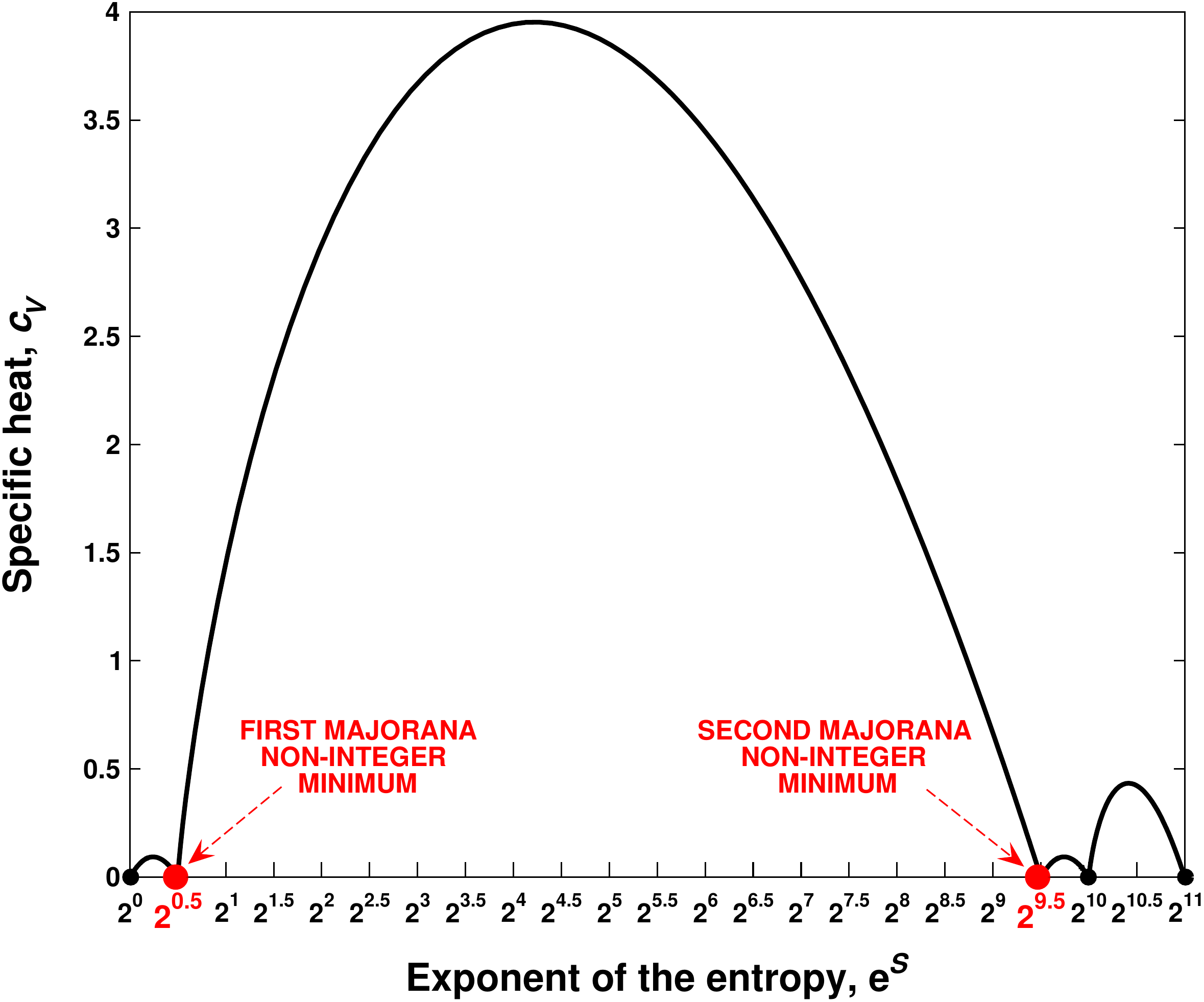}
\caption{\label{figure_9} Specific heat $c_V$ as a function of the exponent of
  the tunneling entropy, $\exp(S)$, for the case $N_\text{TS}=10$,
  $\xi/\Gamma=10^{-7}$. The other parameters have the same values as in
  Fig. \ref{figure_2} where now $\eta$ should be understood as the total
  tunneling strength between the quantum dot and topological
  superconductors. As expected, in addition to the standard minima at $2^0$,
  $2^{10}$ and $2^{11}$ there appear two additional Majorana non-integer minima at
  $2^{0.5}$ and $2^{9.5}$.}
\end{figure}

We further calculate the specific heat $c_V=T\partial S/\partial T$ which is
shown in Fig. \ref{figure_6}. In the absence of the Majorana fermions,
Fig. \ref{figure_6} a), the specific heat has a single peak corresponding to the
transition between the two integer plateaus of the entropy. This peak remains
when the quantum dot is coupled to a topological superconductor,
Fig. \ref{figure_6} b). For the case of the strongly overlapping Majorana
modes (the black curve) there are two standard peaks corresponding to the
transitions between the integer plateaus of the entropy. For the weakly
overlapping Majorana bound states there appears the third peak in the specific
heat. This Majorana peak is induced by the formation of the Majorana plateau
in entropy and represents the first low temperature peak of the specific heat.

The specific heat is a physical observable which can be experimentally
measured. The temperature dependence of the entropy can be recovered from the
temperature dependence of $c_V$. The specific heat then may be plotted versus
the exponent of the entropy, as it is done in Fig. \ref{figure_7}. This
representation is very instructive. In systems with only normal fermions,
Fig. \ref{figure_7} a), the specific heat has standard minima only at
integers. However, when the Majorana fermions strongly govern the low energy
physics of the quantum dot, the specific heat acquires an additional Majorana
minimum at a non-integer number, Fig. \ref{figure_7} b), equal to
$\sqrt{2}$. This non-integer minimum of the specific heat disappears when the
two Majorana modes strongly overlap as it is demonstrated in
Fig. \ref{figure_7} c).

In spite of the fact that the Majorana plateau is small (of the order of
$k_\text{B}$ in SI units) it can be recovered from the specific heat already
in modern experiments and from values of the specific heat which are even
smaller than the ones predicted in Fig. \ref{figure_7}. Indeed, in experiments
of Ref. \cite{Liang_2015} a very smooth entropy curve below and above
$\ln(2)/2$ is recovered from the experimental measurements of the specific
heat which is the quantity used in the integration. This definitely means that
already nowadays high precision measurements of $c_V$ are possible and may
allow to resolve the Majorana plateau and one naturally expects that in the
near future experiments will become even more precise.
\begin{figure}
\includegraphics[width=8.0 cm]{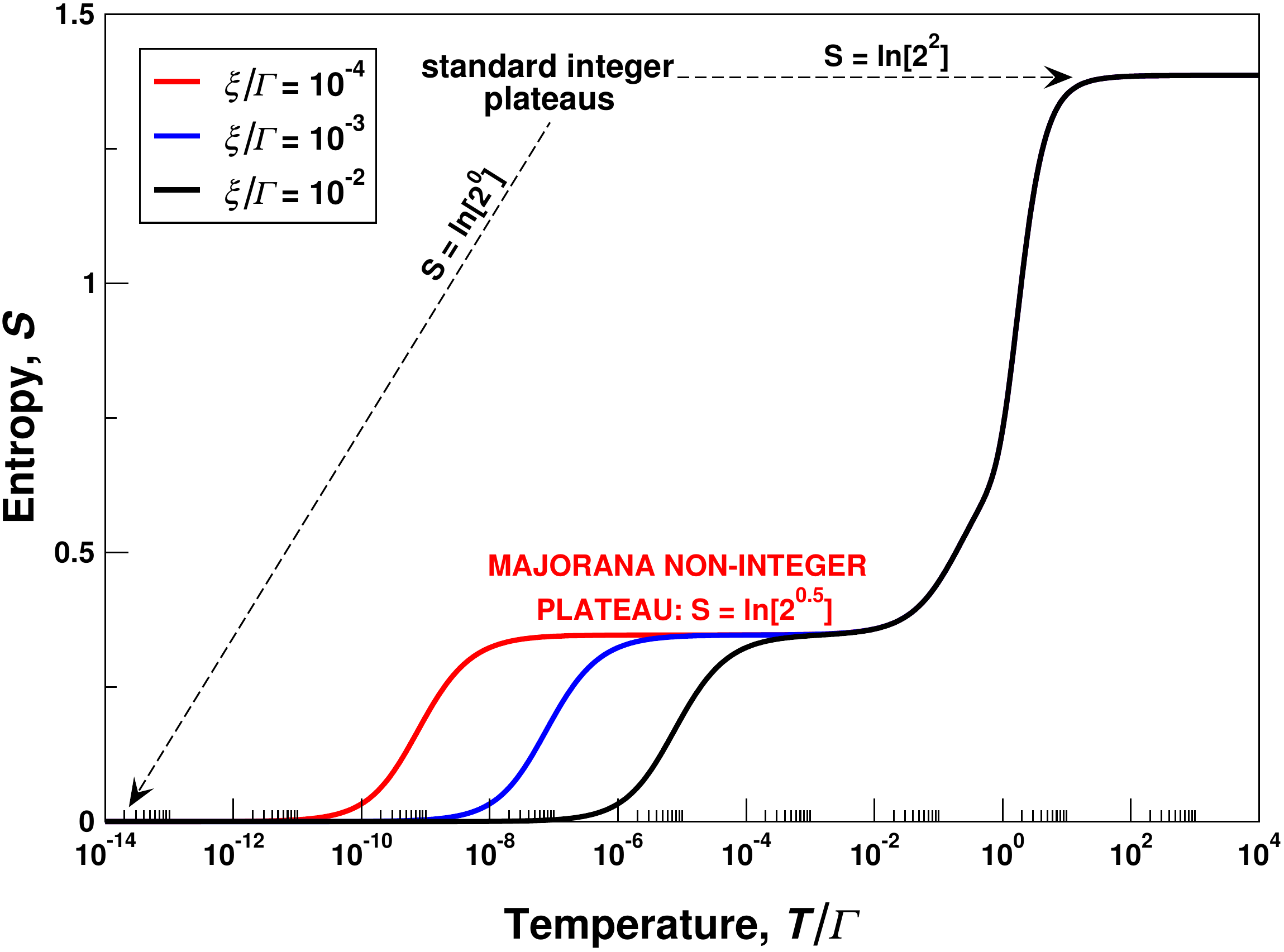}
\caption{\label{figure_10} Tunneling entropy $S$ as a function of the
  temperature $T/\Gamma$ for different values of the overlap energy
  $\xi/\Gamma$ of the Majorana bound states. The other parameters are
  $\epsilon_d/\Gamma=-1.0$, $\eta/\Gamma=2.0$. The standard integer plateau
  with $\exp(S)=2^1$ is washed out by the significantly expanding non-integer
  Majorana plateau with $\exp(S)=\sqrt{2}$ which is visible already at
  $T/\Gamma=5\cdot 10^{-3}$.}
\end{figure}
\begin{figure}
\includegraphics[width=8.0 cm]{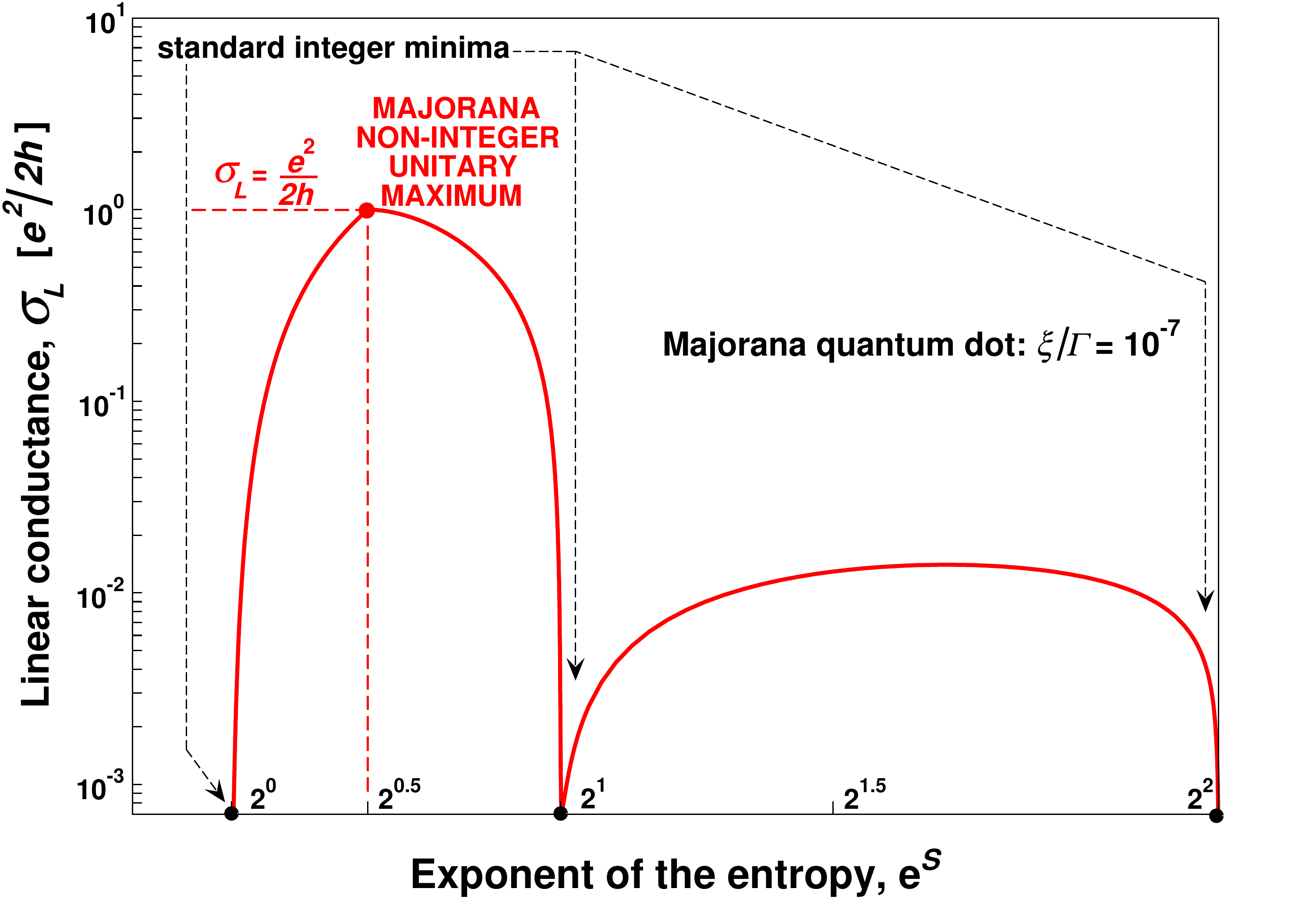}
\caption{\label{figure_11} Linear conductance $\sigma_L$ as a function of the
  exponent of the tunneling entropy, $\exp(S)$. In this representation the
  linear conductance has the standard minima at the integer values $2^0$,
  $2^1$ and $2^2$. The two Majorana modes overlap weakly,
  $\xi/\Gamma=10^{-7}$, and this results in a Majorana unitary maximum,
  $\sigma_L=e^2/2h$, at the non-integer value $\sqrt{2}$. The other parameters
  are as in Fig. \ref{figure_2}.}
\end{figure}

Moreover, the Majorana non-integer entropy plateaus can be experimentally
recovered using several topological superconductors $N_\text{TS}>1$ as
discussed above in connection with Figs. \ref{figure_4} and
\ref{figure_5}. Indeed, as it is obvious from Figs. \ref{figure_4} and
\ref{figure_5}, the specific heat coming from the transition between the two
non-integer Majorana plateaus will grow with $N_\text{TS}$. Therefore, it will
be even easier to experimentally detect the corresponding peak in the specific
heat and its two non-integer edges as it is shown in Figs. \ref{figure_8} and
\ref{figure_9} for $N_\text{TS}=4$ and $N_\text{TS}=10$, respectively.

Concerning the temperatures at which the Majorana non-integer plateau can be
experimentally observed we find that it may be visible already at
$T/\Gamma=5\cdot 10^{-3}$ as shown in Fig. \ref{figure_10}. This temperature
is well within the range of modern experiments. Indeed, for
$\Gamma\approx\Delta$ ($\Delta$ is the induced superconducting gap) one
obtains $T\approx 1\,K$ for $\Delta=15\,meV$ from the experiments of
Ref. \cite{Wang_2013}. The temperatures $T\approx 1\,K$ are easily reachable in
modern physical laboratories. Even with $\Delta=250\,\mu eV$ from the
experiments of Ref. \cite{Mourik_2012} one gets $T\approx 15\,mK$ which is
within modern experiments as well.

Finally, let us consider an application of the thermodynamic analysis to the
transport properties. For example, the linear conductance, $\sigma_L$, which
is a transport property characterized by an equilibrium state of the quantum
dot, may be found as the derivative $\sigma_L=\partial I/\partial V$ of the
current,
\begin{equation}
\begin{split}
&I=-\frac{e\Gamma}{4\pi\hbar^2}\int_{-\infty}^\infty
d\epsilon \,\,\text{Im}[G_{hp}^R(\epsilon)][n_\text{L}(\epsilon)-n_\text{R}(\epsilon)],\\
&n_\text{R,L}(\epsilon)=\frac{1}{\exp[\beta(\epsilon\pm eV/2)]+1},
\end{split}
\label{MW}
\end{equation}
with respect to the voltage $V$ at $V=0$. In Eq. (\ref{MW}) $e$ is the
electronic charge and $n_\text{R,L}(\epsilon)$ are the Fermi-Dirac
distributions in the right and left normal metallic contacts and we have
assumed $\Gamma_\text{L}=\Gamma_\text{R}$ for simplicity. When the two
Majorana bound states in the topological superconductor overlap weakly the
linear conductance at low temperatures reaches the unitary maximum $e^2/2h$,
where $h$ is the Planck's constant. In many recent publications
\cite{Vernek_2014,Cheng_2014,Liu_2015,Tijerina_2015} it is argued that this
unitary maximum is a result of the tunneling coupling of the quantum dot to
the topological superconductor. Our results provide an alternative
thermodynamic explanation based on the structure of the macroscopic state of
the quantum dot. As it is shown in Fig. \ref{figure_11}, the linear conductance
as a function of the exponent of the entropy reaches the Majorana unitary
maximum at the non-integer value $\exp(S)=\sqrt{2}$ which corresponds to the
Majorana plateau in Fig. \ref{figure_2}. This shows that the unitary maximum
of the linear conductance is a consequence of the quantum dot macroscopic
state composed of a non-integer number of microscopic states, corresponding to
the half-fermionic value of the entropy, and that the temperature range of the
linear conductance unitary maximum coincides with the temperature range of the
non-integer Majorana plateau of the tunneling entropy.
\section{Conclusion}\label{Concl}
To conclude, we would like to emphasize two aspects of our work which are
fundamental for the Majorana physics in general but currently not well
explored. Our results demonstrate the fundamental role of the entropy in the
deep physical understanding of macroscopic states of systems coupled to
Majorana bound states and, as a result, in the deep physical understanding of
the behavior of their observables. Therefore, the thermodynamics of such
systems represents on one side an independent research field with its own,
thermodynamic, signatures for Majorana fermions and on the other side it
provides a connection to other fields, such as transport, advancing their
further understanding on a deep fundamental basis.

\section{Acknowledgments}
Support from the DFG under the program SFB 689 is acknowledged.

\end{document}